\documentclass[prx,onecolumn,showpacs,amsmath,amssymb,superscriptaddress]{revtex4-2}
\usepackage{graphicx}
\usepackage{dcolumn}
\usepackage{bm}
\usepackage{hyperref}
\hypersetup{colorlinks,citecolor=blue, filecolor=blue, linkcolor=blue , urlcolor=blue}
\usepackage{ulem}
\usepackage[version=3]{mhchem}
\usepackage{hyperref}
\usepackage{gensymb}
\usepackage[utf8]{inputenc}
\usepackage{textcomp}
\begin{document}
    \title{Origin of metal-insulator transition in rare-earth Nickelates}
	\author{Swagata\,Acharya}
    \email{swagata.acharya@nrel.gov}
    	\affiliation{National Renewable Energy Laboratory, Golden, Colorado 80401}
	\author{Brooks\,Tellekamp}
	\affiliation{National Renewable Energy Laboratory, Golden, Colorado 80401}
	\author{Jerome Jackson}
    \affiliation{Scientific Computing Department, STFC Daresbury Laboratory, Warrington WA4 4AD, United Kingdom}
	\author{Dimitar\,Pashov}
    \affiliation{King's College London, Theory and Simulation of Condensed Matter, The Strand, WC2R 2LS London, UK}
    \author{Jeffrey\, L. Blackburn}
    \affiliation{National Renewable Energy Laboratory, Golden, Colorado 80401}
    \author{Kirstin\,Alberi}
    \affiliation{National Renewable Energy Laboratory, Golden, Colorado 80401}
	\author{Mark\,van Schilfgaarde}
	\affiliation{National Renewable Energy Laboratory, Golden, Colorado 80401}

	\begin{abstract}
        Rare-earth nickelates RNiO$_{3}$ (R=rare-earth element) exhibit three kinds of phase transitions with decreasing temperature: a structural
        transition from a pseudo-cubic to a monoclinic  phase, a metal-insulator transition (MIT), and a magnetic transition from a 
        paramagnetic state to an ordered one.  The first two occur at the same temperature, which has led to a consensus
        that the MIT is driven by lattice distortions.  We show here that the primary driving force for the MIT is
        magnetic; however because of the unusual $d^{7}$ configuration of Ni, additional flexibility in spin
        configurations are also needed which symmetry-lowing structural deformations make possible.  The latter
        enable Ni to disproportionate into two kinds: a high-spin and a low-spin configuration, which allow
        the system to reduce its unfavorable orbital moment and also open a gap.
	\end{abstract}
    
	\maketitle

\section{Introduction}
\label{sec:intro}

Rare-earth nickelates RNiO$_3$ (R = rare-earth element) belong to the ABO$_3$ perovskite family and exhibit a rich phase
diagram encompassing metal-insulator transitions (MIT), magnetic ordering, structural distortions, and even
superconductivity under specific perturbations such as strain or doping. As the ionic radius of the rare-earth element
increases (moving leftward across the lanthanide series), the critical temperature for the MIT decreases, with LuNiO$_3$
exhibiting the highest~\cite{Serrano2022} and PrNiO$_3$ the lowest transition
temperatures~\cite{guo2018antiferromagnetic}. LaNiO$_3$ is a notable exception—it remains metallic down to the lowest
measured temperatures and does not undergo any structural distortion~\cite{Golalikhani2018}. The origin of the MIT has
been a subject of intense debate, with two primary schools of thought: one attributing the transition to structural
distortions, and the other to electronic and magnetic instabilities.

\begin{figure*}
	\includegraphics[scale=0.55]{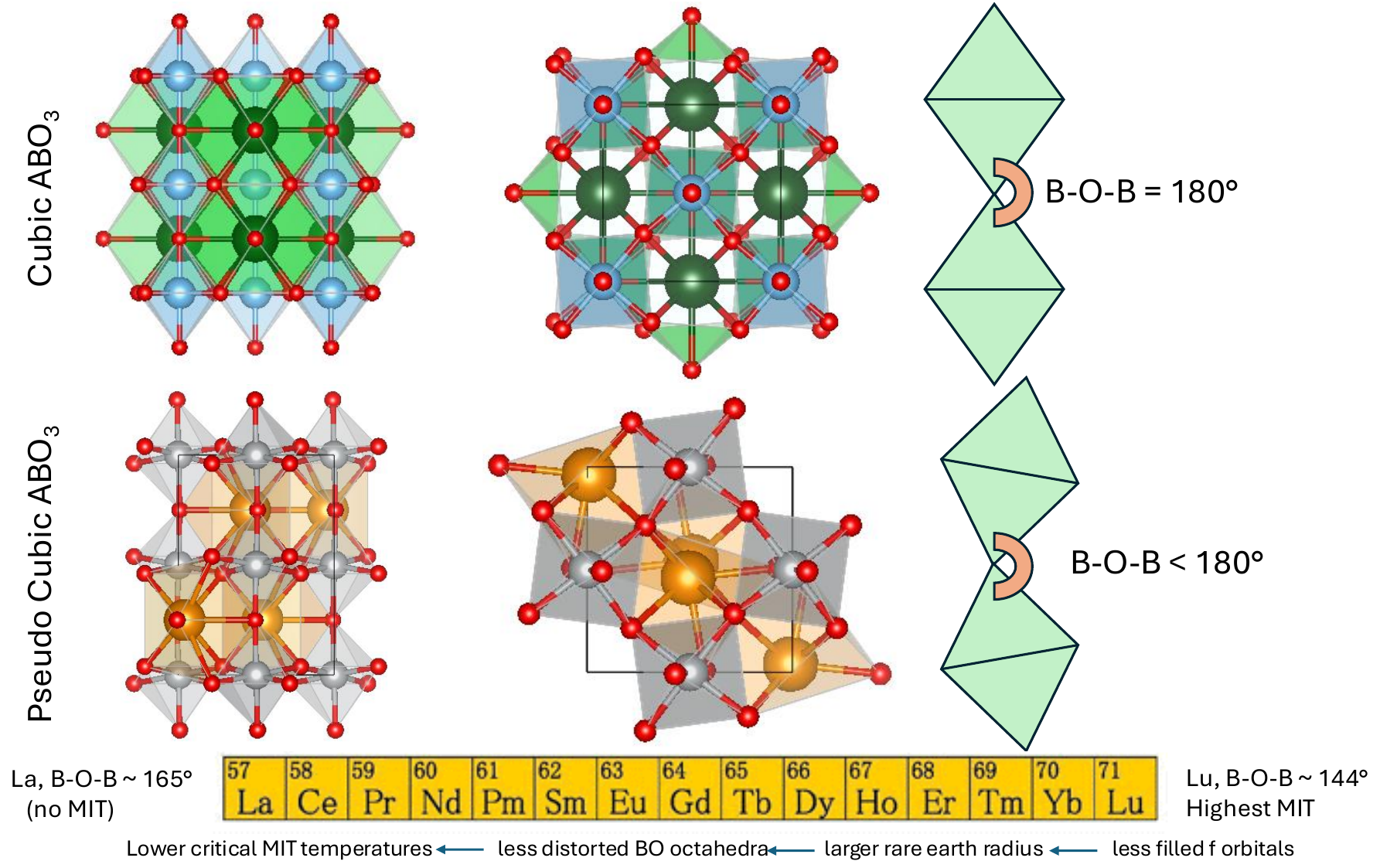}
	\caption{{\bf Nature of the pseudo-cubic distortion in RNiO$_3$:} In RNiO$_3$ compounds, the pseudo-cubic distortion causes the B-O-B bond angle (B is Ni in this case) to deviate from the ideal 180$^\circ$. This distortion is influenced by the size of the rare-earth ion: ions with less-filled $f$-orbitals tend to have a larger ionic radius. As the rare-earth ionic radius increases, the degree of pseudo-cubic distortion decreases. Although there is a correlation between the metal-insulator transition (MIT) temperature and the extent of pseudo-cubic distortion, the orthorhombic \textit{Pbnm} phase alone is not sufficient to support an insulating state. Achieving insulation requires an additional reduction in crystal symmetry beyond the \textit{Pbnm} structure.}
	\label{fig:summary1}
\end{figure*}

At high temperatures ($T > T_{\text{MIT}}$), all RNiO$_3$ compounds (except LaNiO$_3$) adopt an orthorhombic
\textit{Pbnm} structure and exhibit a paramagnetic metallic phase. The degree of distortion from the ideal cubic
perovskite structure in the \textit{Pbnm} is influenced by the size of the rare-earth ion, which affects the Ni–O–Ni
bond angle~\cite{Catalano2018}. The larger the pseudo cubic distortion is, the higher $T_{\text{MIT}}$
(Fig.~\ref{fig:summary1}). However, this structural tuning alone cannot explain the MIT. It is accompanied by a
symmetry-lowering transition from the orthorhombic \textit{Pbnm} to the monoclinic \textit{P2$_1$/n} phase. This
transition entails a breathing-mode distortion, where NiO$_6$ octahedra alternate between expanded and contracted
forms, leading to two inequivalent Ni sites~\cite{Mercy2017,Badrtdinov2021}. First-principles calculations have shown
that this breathing mode is structurally triggered by oxygen octahedral rotations, and its softening is essential for
the MIT~\cite{Mercy2017}. The structural distortion thus appears necessary to stabilize the insulating phase.

However, several studies argue that the MIT is fundamentally electronic or magnetic in origin. Lau and
Millis~\cite{Lau2013} proposed that the transition is driven by magnetism, with structural distortions playing a
secondary, enabling role. Similarly, Badrtdinov \textit{et al.}~\cite{Badrtdinov2021} demonstrated that the
breathing-mode distortion is strongly coupled to the antiferromagnetic order, and that the competition between
nearest-neighbor ferromagnetic and next-nearest-neighbor antiferromagnetic interactions determines the magnetic ground
state. Further support for the electronic origin comes from hybrid functional and DFT+U+V calculations, which show that
both on-site (U) and inter-site (V) Coulomb interactions are required to reproduce the insulating state and the observed
bond disproportionation~\cite{Binci2023}. Without these electronic interactions, the breathing distortion and the band
gap fail to emerge in simulations.

The unique magnetic nature of Ni cannot be overlooked either. Unlike its natural valence of 2$^+$, Ni in these compounds
nominally adopts a 3$^+$ valence state, corresponding to a partially filled magnetic $d^7$
configuration~\cite{Freeland2015}. This configuration allows for multiple spin states and strong hybridization with
oxygen ligands, leading to complex magnetic behavior. The structural transition from \textit{Pbnm} to \textit{P2$_1$/n}
typically takes the system from a paramagnetic metal to a paramagnetic insulator. Only for larger rare-earth radii
(e.g., Nd, Pr) does the MIT coincide with a transition to an antiferromagnetic insulating
phase~\cite{Binci2023}. Regardless of the rare-earth ion, magnetism, whether ordered or disordered, is omnipresent in
both metallic and insulating phases. This raises a fundamental question: what role does magnetism play in enabling or
stabilizing the structural distortion, and how does it influence the magnetic structure of the inequivalent Ni sites in
the monoclinic phase?

This question echoes a broader debate often framed as a ``chicken-and-egg'' problem: is it the magnetism or the structural distortion that drives the MIT in systems where they occur at the same critical temperature? A classical example of this dilemma is found in VO$_2$, where the rutile-to-monoclinic (M1) structural distortion is accompanied by a first-order MIT. Over the decades, extensive research has argued both for an electronically driven MIT via Mott mechanisms~\cite{PhysRevB.78.115103,vo2_paper_ref1,silke_vo2,jan_vo2,PhysRevB.95.035113,Pouget2021,Si2012,Lee2023} and for a lattice-driven transition~\cite{vo2_pump2,vo2_pump3,vo2_pump4,vo2_pump5,hiroi2015structural,del2025structural,weber2020role}.
Recent advances in ultrafast spectroscopy have provided new tools to disentangle these coupled degrees of freedom. By probing systems on femtosecond timescales, it becomes possible to temporally separate the electronic and structural responses to external stimuli. In VO$_2$, such studies have revealed that the electronic transition can precede the structural one, suggesting a decoupling of mechanisms that are otherwise entangled in equilibrium~\cite{Hwang2021,Lee2023,Han2021}. This approach offers a promising pathway to investigate similar questions in RNiO$_3$, where the coupling between spin, lattice, and charge degrees of freedom remains a central puzzle.

The theoretical investigation into this faces its own challenges: it is difficult to construct a theoretical framework that captures the interplay between structural distortions and magnetism across the entire phase diagram in an unbiased manner. This challenge is compounded by the limitations and ambiguities inherent in commonly used computational approaches such as DFT+U and dynamical mean-field theory (DMFT). DFT+U, while widely used, tends to favor magnetic solutions that can artificially open a gap in the electronic structure. This makes it difficult to disentangle whether the gap is a consequence of intrinsic magnetic ordering or an artifact of the method~\cite{Pavarini2017}. Moreover, the inclusion of inter-site Coulomb interactions (V) and Hund's coupling (J) introduces additional parameters that are often empirically chosen, further complicating interpretation~\cite{Binci2023}.

On the other hand, DMFT provides a more sophisticated treatment of local electronic correlations and is particularly
well-suited for capturing Mott physics and Kondo-like behavior. However, it suffers from its own set of challenges,
including ambiguities in the choice of Hubbard parameters, double-counting corrections, and the embedding of the
impurity problem into the lattice~\cite{Chen2022, Carta2025}. Furthermore, conventional single-site DMFT is exact only
in the limit of local interactions and fails to capture non-local correlations that are essential in systems like
RNiO$_3$, where the MIT is accompanied by a symmetry-lowering distortion involving two inequivalent Ni sites. The nature
of the monoclinic \textit{P2$_1$/n} distortion, which introduces Ni$_1$ and Ni$_2$ sublattices with distinct local
environments, suggests that the minimal model capable of describing the insulating phase must include at least a
two-site Ni cluster and their associated ligand states. This necessitates the use of cluster-DMFT approaches, which can
capture inter-site correlations and the cooperative nature of the breathing-mode distortion~\cite{Fursich2019}. While
cluster DMFT is an attractive tool for NdNiO$_3$ and related compounds to explore the insulating nature of the
low-temperature phase,~\cite{Li2019} systematic disentangling of the causal relationship between structural symmetry
breaking, magnetism, and the MIT remains a challenge within such framework.

\begin{figure*}
        \includegraphics[scale=0.40]{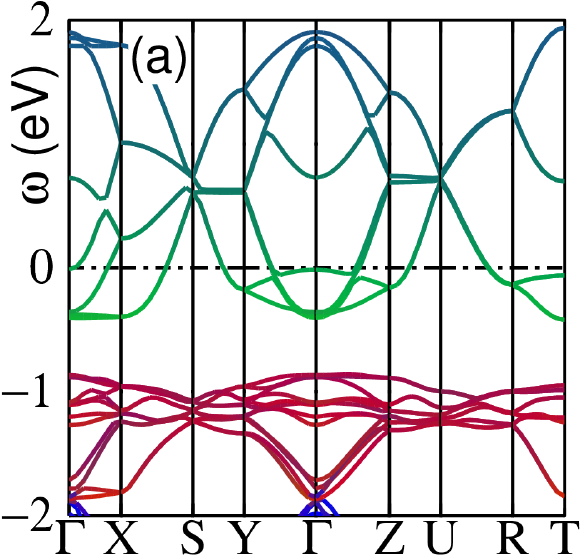}
        \includegraphics[scale=0.40,trim=-1cm 0 0 0]{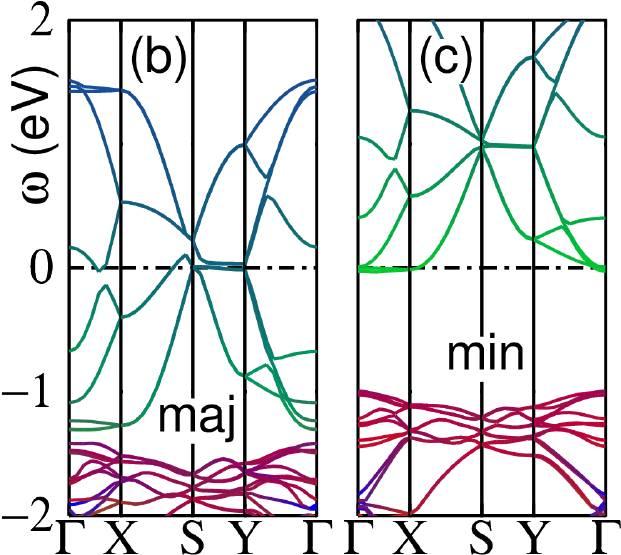}
        \includegraphics[scale=0.35,trim=0 1cm 0 0]{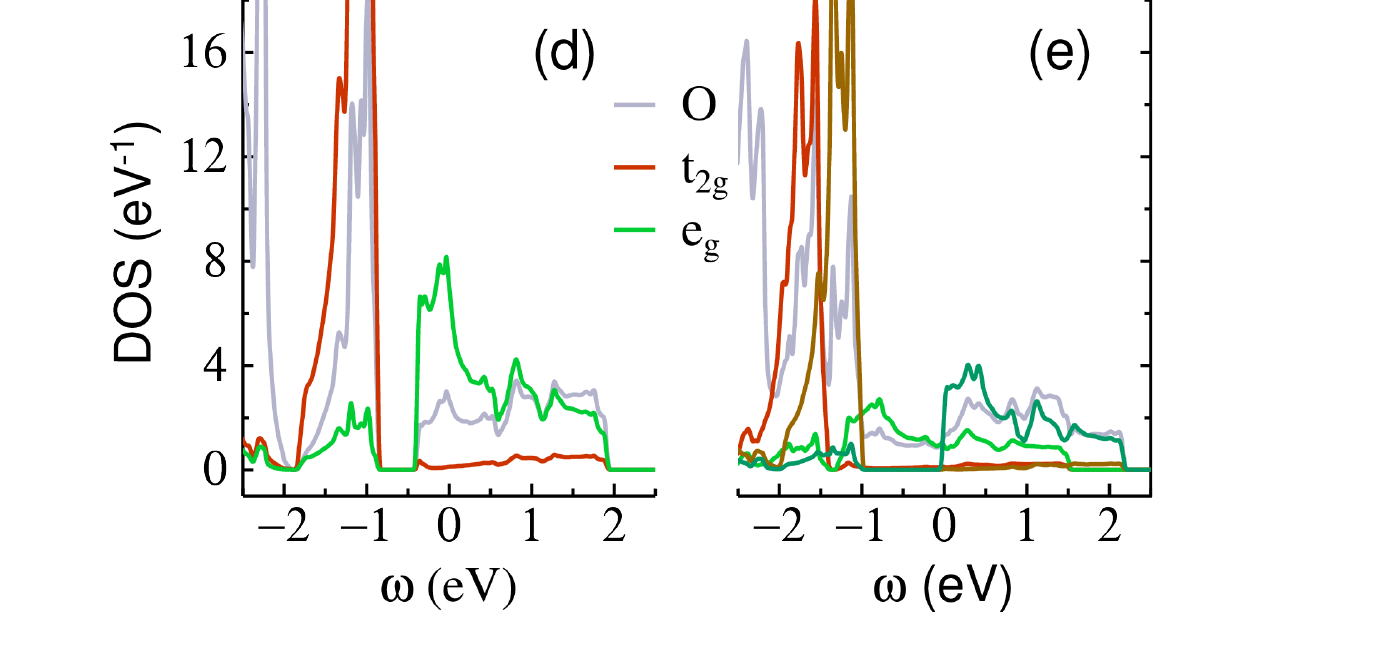}
        \caption{NdNiO$_{3}$ in the pseudocubic Pbnm phase.
        (a) energy bands in the nonmagnetic case. The $\Gamma$-S-Y lines deviate from mirror images of the $\Gamma$-Z-U lines because of deviations from cubic symmetry.
        Colors represent the following projections:  Ni-$t_{2g}$ (red); Ni-$e_{g}$ (green); O-$p$ (blue); thus the $e_{g}$ fall above the $t_{2g}$.
        (b,c) energy bands in the ferromagnetic case for (majority, minority) spins.
        The bands are similar except $t_{2g}$ and $e_{g}$ are spin split, enabling a gap to form in the minority channel.
        (d) Nonmagnetic density-of-states, showing the non-negligible amount of O character around $E_{F}$=0.
        (e) FM density-of-states, showing how the majority and minority Ni d states are spin split by $~\sim$1\,eV.
        To distinguish spins, colors are modified to: $t_{2g}$ (rust=majority, light-red=minority); $e_{g}$ (green=majority, olive=minority);
        O-$p$ (cornflower blue).
 }
        \label{fig:pbnm:nm}
\end{figure*}

To rigorously describe the electronic structure of the entire series from first principles, we employ a self-consistent
implementation of many-body perturbation theory (MBPT) built into the Questaal
package~\cite{questaal_paper,questaal_web}. A distinguishing feature of Questaal is its quasiparticle self-consistent
form of \textit{GW}, QS\textit{GW}. \textit{GW} is usually implemented as an extension to density-functional theory
(DFT), i.e. \textit{G} and \textit{W} are generated from DFT. QS\textit{GW}~\cite{Faleev04} may be thought of as an
optimized form of the \textit{GW} approximation of Hedin~\cite{mark06qsgw,Kotani07,Beigi17}.  Self-consistency removes
dependence on the starting point~\cite{mark06qsgw,Kotani07} and also makes it possible to more systematically and
reliably predict physical observables, especially those sensitive to self-consistency such as the magnetic
moment~\cite{Faleev04} and response functions~\cite{Acharya21a}. We address the MIT across the entire series by
performing electronic structure calculations for seven candidate systems (R=Lu, Yb, Tm, Er, Ho, Nd, Pr). We find that the
QS\textit{GW} finds a metallic solution in the \textit{Pbnm} phase and an insulating solution in the monoclinic phase,
in agreement with experiments, for all candidate systems. Further, by artificially biasing a spin-disproportionate
solution in the \textit{Pbnm} phase and a spin-equivalent solution in the \textit{P2$_1$/n} phase we disentangle the
role of structure and magnetism in driving the MIT. One key achievement of our work is that this disentanglement could
be achieved without making any model Hamiltonian approximation and without any free parameters.

\section{Results}
\label{sec:results}
\section*{Non-magnetic (nm) electronic structure in $Pbnm$ phase}

We begin by examining the electronic structure of NdNiO$_3$ in its high-symmetry \textit{Pbnm} phase, as a representative example of the RNiO$_3$ family. NdNiO$_3$ is particularly interesting because it features both partially filled Ni-$d$ and Nd-$f$ states. However, the Nd-$f$ orbitals lie well away the Fermi level ($E_F$) and do not contribute to the low-energy physics. Instead, the electronic structure near $E_F$ is dominated by hybridized Ni-$d$ and O-$p$ states. In the \textit{Pbnm} phase, all NiO$_6$ octahedra are crystallographically equivalent, and consequently, all Ni atoms exhibit identical spin and orbital moments. Our calculations reveal an unquenched orbital moment of approximately 0.10~$\mu_B$ per Ni atom. This is notably larger than the orbital moment observed in cubic NiO, which is about 0.03~$\mu_B$~\cite{acharya2023theory}. The enhanced orbital moment in NdNiO$_3$ reflects the distinct electronic environment of Ni in RNiO$_3$ compared to NiO.

The key differences arise from both the nominal valence and the crystal field environment. In NiO, Ni is in a formal $d^8$ configuration, stabilized by a nearly perfect cubic crystal field that splits the $d$ orbitals into a fully filled $t_{2g}^6$ and a half-filled $e_g^2$ manifold. This configuration supports strong electron-electron repulsion, leading to a Mott insulating state. In this case, the energy cost of double occupancy on a Ni site outweighs the kinetic energy gain from electron hopping, thereby localizing the electrons and opening a correlation-driven gap. Importantly, NiO remains insulating in both its magnetically ordered and disordered phases, underscoring the purely electronic nature of its insulating behavior. In contrast, Ni in RNiO$_3$ is nominally in a $d^7$ configuration, and the lower symmetry of the \textit{Pbnm} phase introduces additional complexity through octahedral tilts and distortions. These structural features, combined with strong hybridization with oxygen, give rise to a more intricate interplay between lattice, spin, and orbital degrees of freedom, which is central to understanding the MIT in these materials.

\section*{Constrained Spin-disproportionated electronic structure in $Pbnm$ phase}

\begin{figure*}
	\includegraphics[scale=0.60]{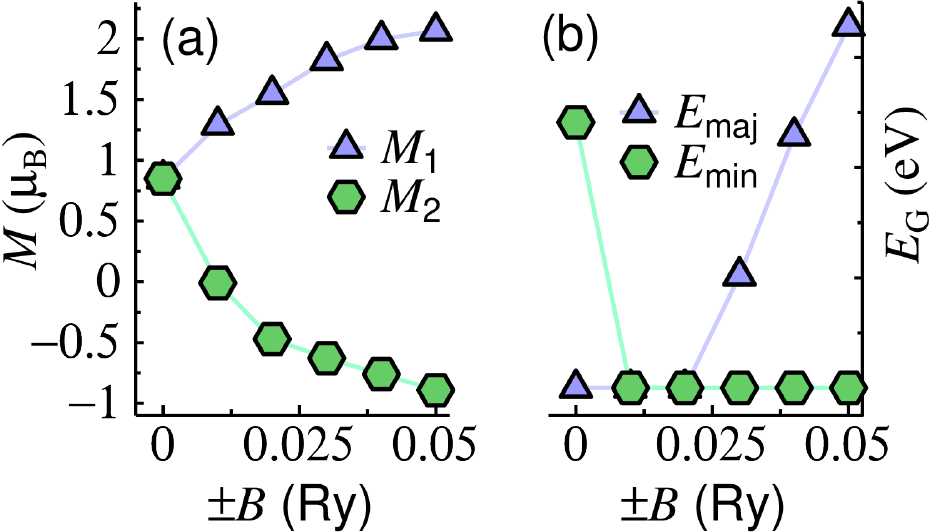} \hspace{1cm}
	\includegraphics[scale=0.60]{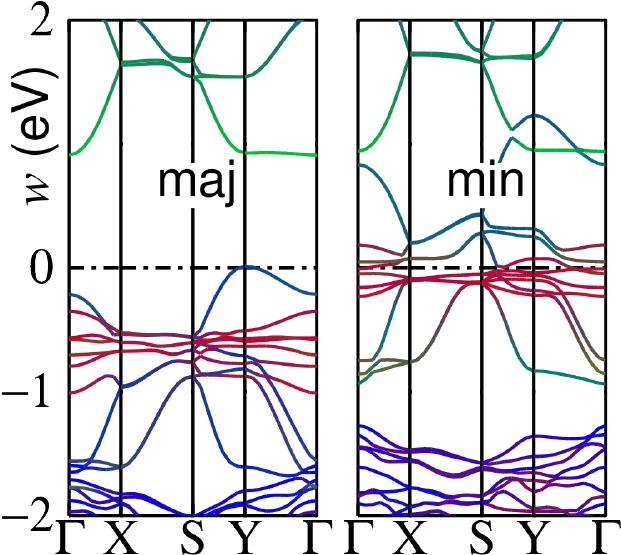}
	\caption{Local magnetic moments (a) and bandgaps (b) of Pbnm NdNiO$_{3}$ in the presence of a site-local magnetic field $\pm{\textit{B}}$.
        Right panels show energy bands for ${\textit{B}}=0.04$\,Ry.}
	\label{fig:pbnm:ferri}
\end{figure*}

To disentangle these mechanisms, we explore further different magnetic configurations of the \textit{Pbnm} phase. A ferromagnetic solution of Pbnm can also be stabilized (Fig.~\ref{fig:pbnm:nm}(b,c)) with self-consistent moment 0.85~$\mu_B$ (all Ni sites are equivalent).  The band structure is similar to the nonmagnetic one (Fig.~\ref{fig:pbnm:nm}(a)) except the Ni d states are spin split.  As in the nm case states near $E_{F}$ are predominantly of $e_{g}$ character.  Spin splitting enables a gap of $\sim$1\,eV to open in the minority channel, but not the majority.  This offers the first hint that the MIT is driven by magnetism, and that the lattice distortion plays a secondary role.  To open the gap, some form of antiferromagnetism is needed. To see this in more detail, consider adding a site-specific external Zeeman field ${\textit{B}}$, driving the system ferrimagnetic.  We do this by retaining only the mirror symmetry in \textit{z}, which splits the four equivalent Ni into two inequivalent pairs.  We apply $B$ parallel or antiparallel to $\textbf{\textit{z}}$ on the respective equivalent sites, thus splitting moments into $M_1$ and $M_2$, then carry the QS\textit{GW} cycle to self-consistency for each ${\textit{B}}$, and trace the evolution of $M_1$, $M_2$, and bandgaps with ${\textit{B}}$.  The result is shown in Fig.~\ref{fig:pbnm:ferri}. Remarkably, the minority-spin gap closes with increasing ${\textit{B}}$, while the majority-spin gap opens, with neither channel having a gap at intermediate ${\textit{B}}$.  As in the FM case, $e_{g}$ falls above $t_{2g}$ but a gap forms in a single channel only.  (The matter is further complicated by the presence of states with O character.)  This suggests that magnetism is the driving force for the MIT, but that Pbnm lacks sufficient configurational freedom to open the gap.  For the monoclinic P2$_{1}$/n phase presented later, enough configurational freedom is available to open a gap in both channels, but in stark contrast to NiO where the spin configuration is of secondary importance, here the gap and local moments are closely connected.  Thus the P2$_{1}$/n distortion provides enough flexibility to enable a gap to open and reduce the orbital moment.

\begin{figure*}
	\includegraphics[scale=0.5]{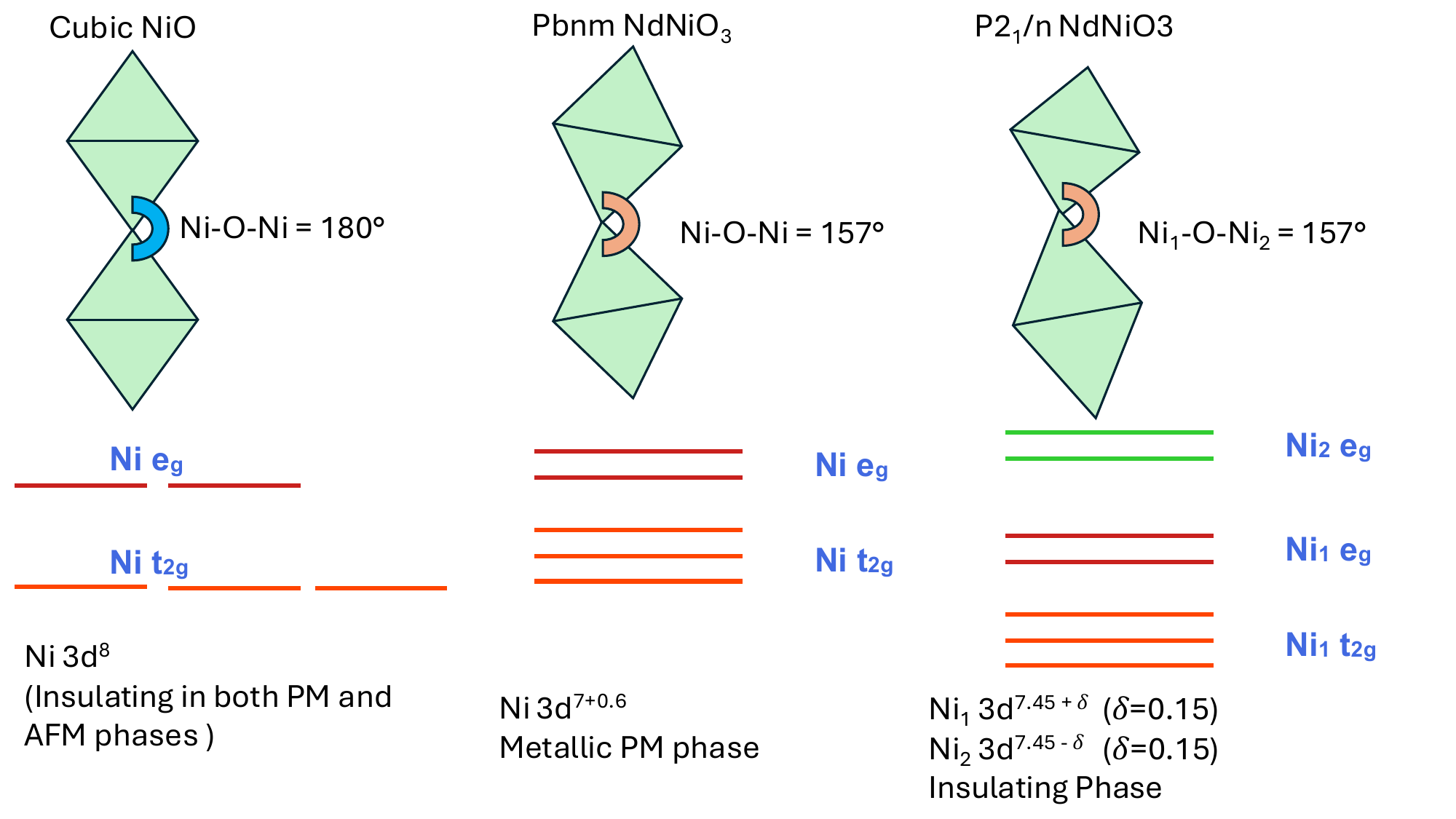}
	\caption{{\bf Schematic representation of cubic, pseudo-cubic \textit{Pbnm} and \textit{P2$_{1}$/n} distortions:} NiO appears in cubic crystal field and is insulating in both the paramagnetic and antiferromagnetic phases. NdNiO$_{3}$ is a paramagnetic metal in the \textit{Pbnm} phase and only becomes insulating in the \textit{P2$_{1}$/n} phase with a concomitant charge, spin and bond disproportionation. Note that the Ni-O-Ni bond angle does not change on average due to the \textit{Pbnm} to \textit{P2$_{1}$/n} distortion. The corresponding crystal field diagrams are also shown for all cases. The insulating gap in \textit{P2$_{1}$/n} phase opens between the e$_{g}$ states of the two distinct Ni sites, while in NiO the band gaps opens at a single Ni site.} 
	\label{fig:summary2}
\end{figure*}

\section*{Antiferromagnetic electronic structure in P2$_{1}$/n phase}

To further understand the insulating state in RNiO$_3$, we simulate the antiferromagnetic (AFM) phase in the low-symmetry monoclinic \textit{P2$_1$/n} structure. Capturing the experimentally observed AFM ordering vector $\left(\frac{1}{2}, 0, \frac{1}{2}\right)$ requires doubling the \textit{Pbnm} unit cell. This doubling naturally accommodates the breathing-mode distortion, which differentiates the NiO$_6$ octahedra into two distinct types: Ni$_1$ and Ni$_2$. Ni$_1$ resides in the larger octahedra with longer Ni–O bond lengths, while Ni$_2$ occupies the smaller octahedra with shorter bonds. The longer Ni–O bonds in Ni$_1$ reduce the hybridization between Ni-$d$ and O-$p$ states, enhancing electronic correlations and favoring a high-spin configuration. In contrast, Ni$_2$ experiences stronger ligand hybridization, stabilizing a low-spin state. This spin-state disproportionation is a hallmark of the insulating phase and is accompanied by charge and orbital moment disproportionation.

Our calculations show that the Nd-$f$ states (and those of other rare-earth ions, except Y where unoccupied $f$ states lie at $E_F + 6$ eV) lie far from the Fermi level, typically a few above and below, and do not contribute to the low-energy physics. All seven RNiO$_3$ compounds studied exhibit indirect band gaps in the AFM phase. For NdNiO$_3$, the valence band maximum (VBM) lies between $E_F$ and $E_F - 0.5$ eV and is primarily composed of Ni$_1$ $e_g$ orbitals ($m_l = -2, 0$), while the conduction band minimum (CBM) originates from the same orbitals on Ni$_2$. This orbital-selective gap is a clear signature of a magnetic gap opening between inequivalent Ni sites~\ref{fig:summary2}.

\begin{table*}[h]
	\begin{center}

		            \begin{tabular}{| c | c| c| c |}
			                \hline
			               & $\left<\mathrm{Ni}_1{-}O\right>-\left<\mathrm{Ni}_2{-}O\right>$, \AA &  AFM gap in meV  & AFM $\left<S\right>$,Ni$_{1}$,Ni$_2$, $\mu_{B}$ \\
			                Y  & 0.071   & 700 &  1.21,0\\
			                Nd & 0.071  & 750 & 1.225,0 \\
			                Ho & 0.08 & 987  & 1.352,0 \\
			                Er & 0.08  &  821 & 1.294,0 \\
			                Tm & 0.094  & 988 & 1.358,0 \\
							Lu & 0.084 & 905 &  1.34,0\\
			               \hline
			            \end{tabular}
		\caption{Theoretically observed bond and spin disproportionation and their relation with computed band gaps across the series.}
		\label{tab:projection}
	\end{center}
\end{table*}
A significant quenching of the orbital moment is observed in the monoclinic phase compared to the high-symmetry \textit{Pbnm} phase. Ni$_1$ retains an orbital moment of 0.074~$\mu_B$ and a spin moment of 1.3~$\mu_B$, while Ni$_2$ exhibits complete quenching of both spin and orbital moments. This remarkable asymmetry is consistent across all RNiO$_3$ compounds studied, indicating that half of the Ni sites in the monoclinic phase become effectively non-magnetic (see Table~\ref{tab:projection}). The projected density of states reveals that Ni$_1$ $t_{2g}$ states lie approximately 1 eV below $E_F$, while its $e_g$ states form the top of the valence band. Conversely, Ni$_2$ $e_g$ states define the bottom of the conduction band. This band alignment is independent of whether FM, PM or AFM simulations are performed. 

A Mulliken charge analysis reveals a real-space charge disproportionation: Ni$_1$ exhibits a $d^{7.6}$ configuration (4.5 electrons in the spin-up channel and 3.1 in spin-down), while Ni$_2$ has a $d^{7.3}$ configuration (3.55 electrons in each spin channel). This corresponds to a charge modulation of $d^{7.45 \pm \delta}$ with $\delta \approx 0.15$. Thus, the monoclinic distortion not only induces bond disproportionation but also drives charge, spin, and orbital moment disproportionation. This results in the spatial separation of two distinct Ni-$d$ spin multiplets across the lattice. In the high-symmetry \textit{Pbnm} phase, these multiplets are dynamically fluctuating and spatially equivalent, leading to a correlated metallic state. The structural distortion in the \textit{P2$_1$/n} phase breaks this symmetry, enabling the system to stabilize these multiplets in real space and open a robust insulating gap. Consequently, the band gap in the AFM insulating phase is fundamentally determined by the energy separation between the high-spin Ni$_1$ and low-spin Ni$_2$ $e_g$ states.

\begin{figure}
        \includegraphics[scale=0.5,angle=0 ]{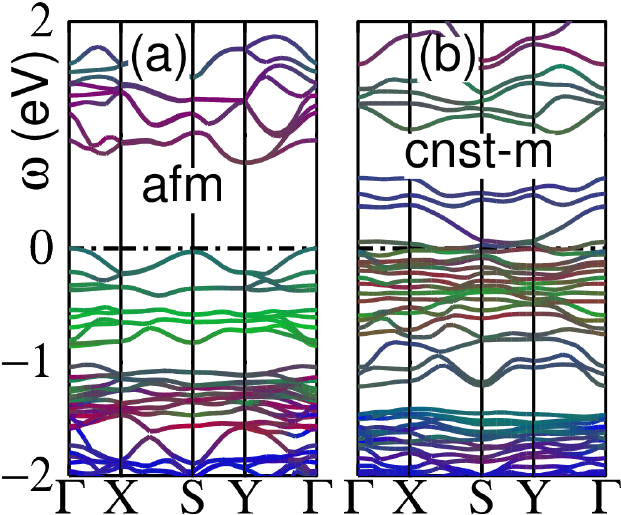}
        \includegraphics[scale=0.5,angle=0,trim=0 1cm 0 0]{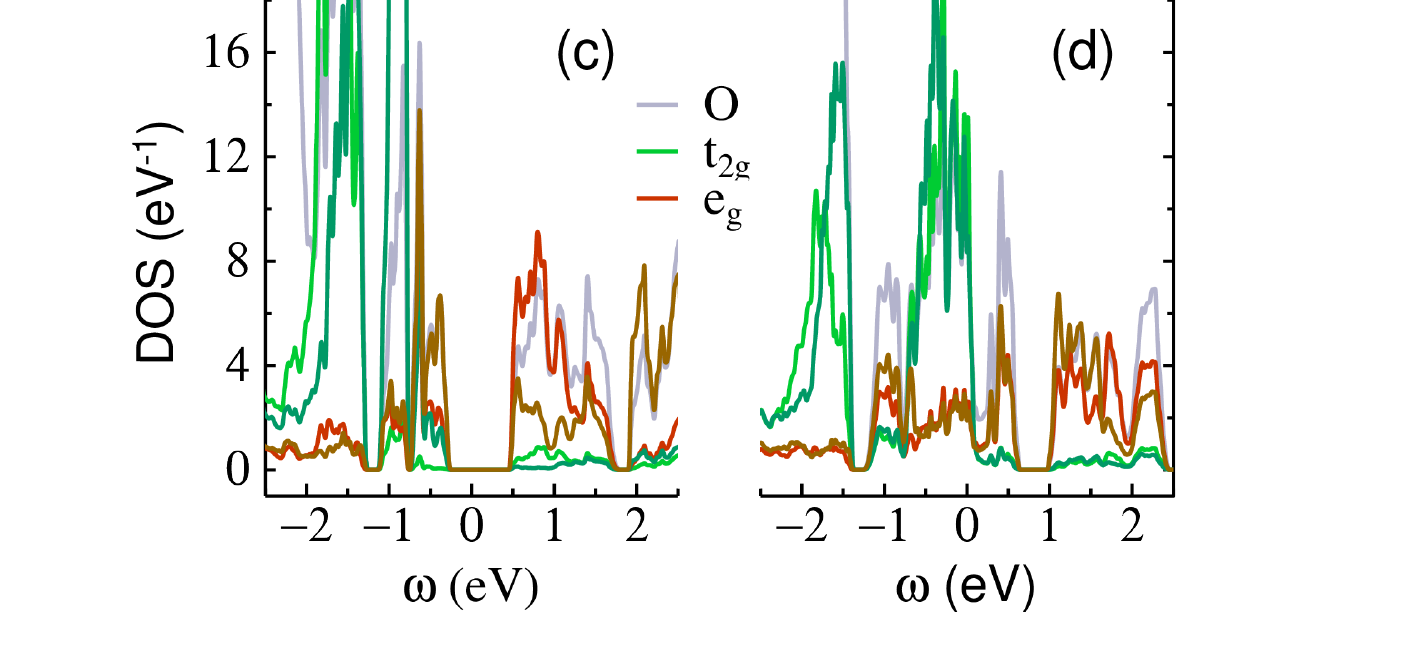} 
        \caption{{\bf AFM NdNiO$_{3}$ in P2$_{1}$/n phase}: (a) Energy bands in AFM
            phase, and (b) the AFM phase with additional constraining fields as described in the text.  (In both cases up-
            and down- bands are nearly equivalent.)  Colors represent the following projections: Ni$_{1}$ (red);
            Ni$_{2}$ (green); O-$p$ (blue).  In the unconstrained case, Ni$_{1}$ and Ni$_{2}$ have local moments 
            ${\pm}1.22\,\mu_B$ and 0, respectively.  In the constrained case, Ni$_{1}$ and Ni$_{2}$ all have approximate
            moments ${\pm}1.2\,\mu_B$.  (c,d) corresponding DOS.  Colors represent the following projections: $e_{g}$
            (rust=Ni$_{1}$, bright-red=Ni$_{2}$); $t_{2g}$ (olive=Ni$_{1}$, bright-green=Ni$_{2}$); O-$p$ (cornflower blue).
}
 	\label{fig:p21nafmdos}
\end{figure}

\section*{Constrained Spin-equivalent electronic structure in P2$_{1}$/n phase}

To show more explicitly that the MIT is a consequence of the bond-disproportionation achieved through unique
combinations of Ni magnetic states, we perform a another calculation, now adding site-local Zeeman fields constraining
all Ni moments to similar absolute values.  Two kinds of fields are added, one on Ni$_{1}$ and another on Ni$_{2}$, to
constrain the local moments on all Ni sites to be ${\pm}1.2\pm{0.2}$\,eV. (There is a small fluctuation in each moment
owing to absence of symmetry.)  The lattice is kept fixed at the breathing-mode distortion and the QS\textit{GW} cycle
is carried through to self-consistency.  Energy band structures for the two cases are compared in
Fig~\ref{fig:p21nafmdos}(a,b).  Fig.~\ref{fig:p21nafmdos}(c,d) show the corresponding Mulliken DOS resolved into
$t_{2g}$ and $e_{g}$ components for Ni$_{1}$ and Ni$_{2}$, and also O $p$ character.  Two key features stand out: First,
that the occupied $t_{2g}$ and $e_{g}$ states split into two peaks, pushing states higher in energy; second that an
O-derived state is split off from the conduction band and bonds with the Ni$_{1}$-$e_{g}$ state, forming a ``midgap''
state around $E_{F}$.  Thus the MIT is formed by a delicate interplay of several interlocking forces:
stabilization through the formation of local spin moments (enabling a gap to open) while also quenching orbital moments
through disproportionation; and also lattice distortions working in tandem with disproportionation to which sweep O $p$
states away from the Fermi level.

\section{Discussion}
The multi-determinantal nature of Mott physics remains a central and unresolved topic in the study of strongly
correlated materials. When combined with structural instabilities, this becomes one of the outstanding problems in
solid-state physics. From that point of view, NiO is relatively simple: it transitions from a paramagnetic insulating to
an antiferromagnetic (AFM) insulating phase without undergoing structural changes. This simplicity makes it a valuable
prototype for studying Mott physics. The classification of systems as truly Mott-like continues to generate debate. NiO
has long been considered a prototypical Mott insulator, often cited as a system that cannot be described within a
conventional Slater framework, but originates from dynamical spin fluctuations.~\cite{Ren2006} Experimental and
theoretical studies have shown that NiO exhibits a charge-transfer gap with strong local Coulomb interactions,
consistent with the Zaanen-Sawatzky-Allen picture~\cite{Wrobel2020}. The Ni$^{2+}$ ion in NiO adopts a $d^8$
configuration with a fully filled $t_{2g}$ shell and two half-filled $e_g$ orbitals, allowing for multiple atomic spin
excitations ($\langle S \rangle = 1$ and $\langle S \rangle = 0$) on individual Ni sites. These dynamic spin
fluctuations are hallmarks of Mott physics.

Our recent many-body perturbative studies~\cite{acharya2023theory}, using a diagrammatic approach that incorporates
long-range Coulomb interactions without frequency-dependent kernels, successfully reproduce the band gap of
NiO. Moreover, the insulating state remains robust against changes in magnetic ordering, indicating that the gap is not
solely tied to magnetic symmetry breaking. This challenges the traditional Mott narrative, yet the nature of the
multi-particle excitations in NiO—such as spinful carriers introduced by weak hole doping and optical access to both
singlet and triplet excitonic states supports its classification as a Mott system. These excitation spectra reflect the
inherently multi-determinantal character of the Ni $d$-manifold, that can be reasonably captured within a single-site
Hubbard model~\cite{Ren2006}.

In contrast, RNiO$_3$ compounds exhibit multi-determinantal physics of a fundamentally different character. Here, the
relevant determinants are spatially distributed across inequivalent Ni sites, making non-local correlations
essential. The insulating state in RNiO$_3$ cannot be characterized in terms of a site-site, frequency-dependent
description, but from real-space disproportionation of charge, spin, and orbital moments between Ni$_1$ and Ni$_2$
sites~\cite{Cui2021}. This spatial separation is stabilized by a monoclinic structural distortion that breaks the
symmetry of the high-temperature metallic phase. A further distinction lies in the orbital moment behavior. In NiO, the
$d^8$ configuration leads to a nearly quenched orbital moment due to high cubic symmetry and strong crystal field
splitting. By contrast, Ni in RNiO$_3$ is in a $d^7$ configuration and exhibits a significantly larger unquenched
orbital moment in the high-symmetry phase. Structural distortions such as Jahn–Teller~\cite{Halcrow2013,Shang2024} or
breathing-mode modulations~\cite{Badrtdinov2021} are known to lift orbital degeneracies and quench orbital moments in a
range of systems, lowering the system's energy. Our theoretical framework shows that while disentangling these coupled
degrees of freedom is challenging, the primary driving force behind the metal-insulator transition in RNiO$_3$ is
magnetic in nature. The breathing-mode distortion plays a secondary, assisting role by enabling spatial separation of
spin multiplets and quenching orbital moments. In summary, NiO exemplifies a Mott insulator governed by local
interactions and dynamic fluctuations, whereas RNiO$_3$ represents a distinct class of correlated insulators where
non-local interactions, structural symmetry breaking, and orbital moment quenching are intricately intertwined. The
minimal Hamiltonian required to describe RNiO$_3$ must include at least two inequivalent Ni sites and their ligand
environments, highlighting the need for theoretical frameworks that go beyond single-site approximations.

\section{Methods: Quasiparticle Self-Consistent \textit{GW}}

The Quasiparticle Self-Consistent \textit{GW} form of \textit{GW}~\cite{mark06qsgw,Kotani07} dramatically improves its
fidelity compared to the usual (DFT-based) \textit{GW}.  Details of the theory and its implementation in
Questaal~\cite{questaal_web} can be found in Refs~\cite{Kotani07,questaal_paper}. QS\textit{GW} largely eliminates the
starting-point dependence (a well known issue for \textit{GW} implementations) and importantly, discrepancies with
experiment become systematic, which clarifies which diagrams are missing.  Also, as a byproduct, QS\textit{GW} yields
the ground state magnetic structure, without reference to DFT--a feature particularly crucial in this study, where the
magnetic ground state is the major question at issue.  As this work shows, the fidelity achievable in QS\textit{GW} and
is essential to distinguish artifacts from lower level approximations.


\section*{Data Availability}

All data will be available from the corresponding authors, SA, on reasonable request. All the input file
structures and the command lines to launch calculations are rigorously explained in the tutorials available on the
Questaal webpage~\cite{questaal_web} \href{https://www.questaal.org/get/}.

\section*{Code Availability}
The source codes for LDA, QS\textit{GW} and QS$G\widehat{W}$ are available from~\cite{questaal_web}  \href{https://www.questaal.org/get/}  under the terms of the AGPLv3 license.

    \bibliographystyle{ieeetr}

\section*{Acknowledgments}
This work was authored in part by the National Renewable Energy Laboratory for the U.S. Department of Energy (DOE) under Contract No. DE-AC36-08GO28308. Funding was provided by the Office of Science, Basic
Energy Sciences, Division of Materials, U.S. Department of Energy.  SA, DP and MvS acknowledge the use of the National Energy Research Scientific Computing Center, under Contract No. DE-AC02-05CH11231 using NERSC award BES-ERCAP0021783, and also acknowledge that a portion of the research was performed using computational resources sponsored by the Department of Energy's Office of Energy Efficiency and Renewable Energy and located at the National Renewable Energy Laboratory, and computational resources provided by the Oakridge leadership Computing Facility. The views expressed in the article do not necessarily represent the views of the DOE or the U.S. Government. The U.S. Government retains and the publisher, by accepting the article for publication, acknowledges that the U.S. Government retains a nonexclusive, paid-up, irrevocable, worldwide license to publish or reproduce the published form of this work, or allow others to do so, for U.S. Government purposes.

\section*{Author Contributions}
SA conceived the work and carried out the calculations and drafted the paper. MvS and DP provided the necessary software
and calculation support. BT, KA and JB provided experimental insights. All authors have contributed to the writing of the paper and the analysis of the data.

\section*{Competing interests}
The authors declare no competing financial or non-financial interests.
\section*{Correspondence}
All correspondence and additional code and data requests should be made to SA.

\end{document}